\documentclass{amsart}
\usepackage{amsmath}
\usepackage{amsfonts}

\newtheorem{theorem}{Theorem}
\theoremstyle{plain}

\newtheorem{definition}{Definition}

\newtheorem{lemma}{Lemma}

\newtheorem{remark}{Remark}

\numberwithin{equation}{section}
\input{tcilatex}

\begin{document}
\title[Equation from theory of the flows on networks]{On some nonlinear
equation from theory of the flows on networks }
\author{Kamal N. Soltanov}
\address{Institute of Math. and Mech. Nat. Acad. of Sci. , Baku, AZERBAIJAN;
Dep. of Math., Fac. of Sci., Hacettepe University, Ankara, TURKEY}
\email{soltanov@hacettepe.edu.tr ; sultan\_kamal@hotmail.com}
\urladdr{}
\date{}
\subjclass[2010]{ Primary 35B30, 35L65, 35R02; Secondary 34A15, 34H05, 37C60}
\keywords{Nonlinear hyperbolic equation, solvability, behaviour}

\begin{abstract}
Here we study the nonlinear hyperbolic equations of the type of equations
from theory of flows on networks, for which we prove the solvability theorem
under the appropriate conditions and also investigate the behaviour of the
solution.
\end{abstract}

\maketitle

\section{Introduction}

In this article we study one class of nonlinear hyperbolic equations, that
in the one space dimension case, we can formulate in the form 
\begin{equation}
u_{tt}-\left( f\left( u\right) u_{x}\right) _{x}=g\left( u\right) ,\quad
\left( t,x\right) \in R_{+}\times \left( 0,l\right) ,\ l>0,  \label{1.1}
\end{equation}%
\begin{equation}
u\left( 0,x\right) =u_{0}\left( x\right) ,\ u_{t}\left( 0,x\right)
=u_{1}\left( x\right) ,\ u\left( t,0\right) =u\left( t,l\right) ,
\label{1.2}
\end{equation}%
where $u_{0}\left( x\right) $, $u_{1}\left( x\right) $ are known functions, $%
f\left( \cdot \right) ,g\left( \cdot \right) :R\longrightarrow R$ are a
continouos functions and $l>0$ is a number. The equation of type (\ref{1.1})
describe mathematical model of the problem from theory of the flow in
networks as is affirmed in articles \cite{a}, \cite{a-r}, \cite{b-c-n}, \cite%
{b-c-g-h-p}, \cite{c-t}, \cite{c-g-p}, \cite{g-p}, \cite{h-k-p}, \cite{h-r}
(e. g. Aw-Rascle equations, Antman--Cosserat model, etc.). As in the survey 
\cite{b-c-g-h-p} is noted such a study can find application in accelerating
missiles and space crafts, components of high-speed machinery, manipulator
arm, microelectronic mechanical structures, components of bridges and other
structural elements. Balance laws are hyperbolic partial differential
equations that are commonly used to express the fundamental dynamics of open
conservative systems (e.g. \cite{c-t}). As the survey \cite{b-c-g-h-p}
possess of the sufficiently exact explanations of the significance of
equations of such type therefore we not stop on this theme. It need to note
that most often in these articles in which the being investigated problem
descrebe the hyperbolic equation of second order as mathematical model, then
for investigation the authors reduce it to the system of equations of first
order. As it is explained in the cited above survey on the mathematical
properties of the Antman--Cosserat model are similar to those of the
first-order system associated with the nonlinear wave equation.

This article is organized as follows. In the section 2 we consider the class
of the nonlinear hyperbolic equations of second order of such type that are
arisen in the theory of flows on networks. In the section 3 we investigate
the solvability of the considered problems and in the section 4\ the
behaviour of their solutions.

\section{Formulation of the problem and the approach}

Consider the following problem 
\begin{equation}
u_{tt}-\nabla \cdot \left( f\left( u\right) \nabla u\right) =g\left(
u\right) ,\quad \left( t,x\right) \in \left( 0,T\right) \times \Omega ,\quad
T\in \left( 0,\infty \right)  \label{2.1}
\end{equation}%
\begin{equation}
u\left( 0,x\right) =u_{0}\left( x\right) ,\ u_{t}\left( 0,x\right)
=u_{1}\left( x\right) ,\ u\left\vert \ _{\left( 0,T\right) \times \partial
\Omega }\right. =0,  \label{2.2}
\end{equation}%
where $\Omega \subset R^{n},n\geq 1$ is a bounded domain with sufficiently
smooth boundary $\partial \Omega $, $T>0$ is arbitrary fixed number.

As it is well known $-\Delta $ is a self-adjoint, positive operator densely
defined in a Hilbert space $H\equiv L^{2}\left( \Omega \right) $ (and on $%
H_{0}^{1}\left( \Omega \right) \equiv H_{0}^{1}$ with the norm $\left\Vert
v\right\Vert _{H_{0}^{1}}\equiv \left\Vert \nabla v\right\Vert
_{L^{2}}\equiv \left\Vert \nabla v\right\Vert _{H^{0}}$, see, e. g. \cite%
{l-m} \cite{t}) and $f,g:R\longrightarrow R$ \ are a continuous functions.

For the investigation of this problem we will use the following approach 
\begin{equation}
\left\langle u_{tt}-\Delta F\left( u\right) ,\left( -\Delta \right)
^{-1}u_{t}\right\rangle =\left\langle g\left( u\right) ,\left( -\Delta
\right) ^{-1}u_{t}\right\rangle ,  \label{2.3}
\end{equation}%
here $F\left( r\right) =\underset{0}{\overset{r}{\int }}f\left( s\right) ds$
is a\ monotone function.

In the other words we will understud the solution of this problem in the
following sense

\begin{definition}
\label{D-1}We will call a function%
\begin{equation*}
u\in C^{0}\left( 0,T;L^{p}\left( \Omega \right) \right) \cap C^{2}\left(
0,T;W^{-2,q}\left( \Omega \right) +H^{-1}\left( \Omega \right) \right)
\end{equation*}
weak solution of problem (\ref{2.1}) - (\ref{2.2}) if $u$ satisfies the
following equation 
\begin{equation*}
\left\langle u_{tt}-\Delta F\left( u\right) ,w\right\rangle =\left\langle
g\left( u\right) ,w\right\rangle
\end{equation*}%
locally by a. e. $t\in \left( 0,T\right) $ for any $w\in W^{2,p}\left(
\Omega \right) \cap H_{0}^{1}\left( \Omega \right) $
\end{definition}

Consider the following conditions

1) let $f,g:R\longrightarrow R$ are a continuous functions and there are a
numbers $a_{0},b_{0},d>0$, $a_{1},b_{1}\geq 0$ and $p>2,$ $0\leq 2p_{0}\leq
p $ such that the following inequations 
\begin{equation*}
\left\vert F\left( r\right) \right\vert \leq a_{0}\left\vert r\right\vert
^{p-1}+a_{1}\left\vert r\right\vert ;\quad F\left( r\right) \cdot r\geq \
b_{0}\left\vert r\right\vert ^{p}+b_{1}r^{2};\quad \left\vert g\left(
r\right) \right\vert \leq d\left\vert r\right\vert ^{p_{0}},
\end{equation*}%
hold for any $r\in R$, moreover $g$ is continuous function (for example, $%
f\left( r\right) =k_{0}\left\vert r\right\vert ^{p-2}-k_{1}\left\vert
r\right\vert ^{p_{1}}+k_{2}$, $k_{0}>0,k_{1},k_{2}\geq 0$, $0\leq p_{1}<p-2$%
, moreover $k_{1}=k_{1}\left( k_{0},k_{2}\right) $ and $g\left( r\right)
=d\left\vert r\right\vert ^{p_{0}}$).

It is well known (\cite{g-t}, \cite{l-m}, \cite{s1}]) that under the
conditions of this problem the following problem 
\begin{equation}
-\Delta v=w,\quad x\in \Omega \subset R^{n},\quad v\left\vert \ _{\left(
0,T\right) \times \partial \Omega }\right. =0  \label{2.4}
\end{equation}%
is solvable for any $w\in L^{p}\left( \Omega \right) $, $p>1$ and has unique
solution in $W^{2,p}\left( \Omega \right) \cap W_{0}^{1,p}\left( \Omega
\right) $, i.e. the operator $-\Delta :W^{2,p}\left( \Omega \right) \cap
W_{0}^{1,p}\left( \Omega \right) \longrightarrow L^{p}\left( \Omega \right) $
is the isomorphism. Consequently if to set the denotation $u\equiv -\Delta v$
then of the posed problem one can rewrite in the form 
\begin{equation*}
-\Delta v_{tt}-\nabla \cdot \left( f\left( -\Delta v\right) \nabla \left(
-\Delta v\right) \right) =g\left( -\Delta v\right) ,\quad \left( t,x\right)
\in \left( 0,T\right) \times \Omega ,
\end{equation*}%
or%
\begin{equation*}
-\Delta v_{tt}-\Delta F\left( -\Delta v\right) =g\left( -\Delta v\right)
,\quad \left( t,x\right) \in \left( 0,T\right) \times \Omega ,
\end{equation*}%
\begin{equation*}
-\Delta v\left( 0,x\right) =u_{0}\left( x\right) ,\ -\Delta v_{t}\left(
0,x\right) =u_{1}\left( x\right) ,\ u\left\vert \ _{\left( 0,T\right) \times
\partial \Omega }\right. =0.
\end{equation*}%
So from equation (\ref{2.3}) we get 
\begin{equation*}
\frac{1}{2}\frac{d}{dt}\left\Vert \nabla v_{t}\right\Vert _{2}^{2}+\frac{d}{%
dt}\Phi \left( -\Delta v\right) =\left\langle g\left( -\Delta v\right)
,v_{t}\right\rangle
\end{equation*}%
here $\Phi $ is a nonnegative functional and $\Phi \left( u\right) =\underset%
{0}{\overset{1}{\int }}\left\langle F\left( su\right) ,u\right\rangle ds$.

If to bear in mind of these conditions and condition 1 we get 
\begin{equation*}
\frac{1}{2}\frac{d}{dt}\left\Vert \nabla v_{t}\right\Vert _{2}^{2}+\frac{d}{%
dt}\Phi \left( -\Delta v\right) \leq \frac{1}{2}\left\Vert v_{t}\right\Vert
_{2}^{2}+\frac{1}{2}\left\Vert g\left( -\Delta v\right) \right\Vert
_{2}^{2}\left( t\right) \leq
\end{equation*}%
\begin{equation*}
c\left[ \frac{1}{2}\left\Vert \nabla v_{t}\right\Vert _{2}^{2}+\Phi \left(
-\Delta v\right) \right] +\widehat{d},
\end{equation*}%
here $c>0,\widehat{d}\geq 0$ are constants that independent of $v$. Hence
follows 
\begin{equation*}
\left\Vert \nabla v_{t}\right\Vert _{2}^{2}\left( t\right) +2\Phi \left(
-\Delta v\right) \left( t\right) \leq e^{ct}\left[ \left\Vert \nabla
v_{1}\right\Vert _{2}^{2}+2\Phi \left( u_{0}\right) \right] +\frac{\widehat{d%
}}{c}\left( e^{ct}-1\right) .
\end{equation*}%
by virtue of the Gronwall's lemma.

Thus we get inequalities 
\begin{equation}
\left\Vert \nabla v_{t}\right\Vert _{2}^{2}\left( t\right) \leq e^{cT}\left[
\left\Vert \nabla v_{1}\right\Vert _{2}^{2}+2\Phi \left( u_{0}\right) \right]
+\frac{\widehat{d}}{c}\left( e^{cT}-1\right)  \label{2.5}
\end{equation}%
\begin{equation*}
\Phi \left( -\Delta v\right) \left( t\right) \leq e^{cT}\left[ \left\Vert
\nabla v_{1}\right\Vert _{2}^{2}+2\Phi \left( u_{0}\right) \right] +\frac{%
\widehat{d}}{c}\left( e^{cT}-1\right)
\end{equation*}%
for every fixed $T\in \left( 0,\infty \right) $

It not is difficult to see that if $\widehat{d}=0$ then occurs the
inequation 
\begin{equation*}
\left\Vert \nabla v_{t}\right\Vert _{2}^{2}\left( t\right) +2\Phi \left(
u\right) \left( t\right) \leq e^{cT}\left[ \left\Vert \nabla
v_{1}\right\Vert _{2}^{2}+2\Phi \left( u_{0}\right) \right] .
\end{equation*}

\section{Solvability of problem (\protect\ref{2.1}) - (\protect\ref{2.2})}

For the proof of the solvability of problem (\ref{2.1}) - (\ref{2.2}) we
will use the approach of Galerkin. We need note that under condition 1 on
the function $g$ we not succeeded in proving the solvability of this
problem. For this we assume the following severe constraint instead of the
condition that is dedused in condition 1 on $g$. Let function $%
g:R\longrightarrow R$ is the Lipschitz function, i.e. there is a such number 
$k>0$ that the following inequality 
\begin{equation}
\left\vert g\left( r\right) -g\left( s\right) \right\vert \leq
d_{0}\left\vert r-s\right\vert  \label{3.1}
\end{equation}%
holds for any $r,s\in R$.

Let the system $U\equiv \left\{ w_{j}\left( x\right) \right\} _{j=1}^{\infty
}$ be a total system of the space 
\begin{equation*}
W^{2,p}\left( \Omega \right) \cap H_{0}^{1}\left( \Omega \right)
\end{equation*}%
where $w_{j}\left( x\right) $ be the sufficiently smooth functions. Using
above approach we get the a priori estimations (\ref{2.5}). We will seek out
of the approximative solutions $u_{m}\left( t,x\right) $ in the form 
\begin{equation*}
\left( -\Delta \right) ^{-1}u_{m}\left( t,x\right) =v_{m}\left( t,x\right) =%
\overset{m}{\underset{i=1}{\sum }}c_{i}\left( t\right) w_{i}\left( x\right) 
\text{ or }u_{m}\left( t\right) \in span\left\{ w_{1},...,w_{m}\right\}
\end{equation*}%
as the solutions of the problem locally with respect to $t$, where $%
c_{i}\left( t\right) $ are as the unknown functions that will be defined as
solutions of the following Cauchy problem for system of ODE 
\begin{equation}
\frac{d^{2}}{dt^{2}}\left\langle u_{m},w_{j}\right\rangle -\left\langle
F\left( u_{m}\right) ,\Delta w_{j}\right\rangle =\left\langle g\left(
u_{m}\right) ,w_{j}\right\rangle ,\quad j=1,2,...,m  \label{3.2}
\end{equation}%
\begin{equation*}
u_{m}\left( 0,x\right) =u_{0m}\left( x\right) ,\ u_{tm}\left( 0,x\right)
=u_{1m}\left( x\right) ,
\end{equation*}%
where $u_{0m}$ and $u_{1m}$ are contained in $span\left\{
w_{1},...,w_{m}\right\} $, $m=1,2,...$, moreover 
\begin{equation*}
u_{0m}\longrightarrow u_{0}\quad \text{in \ \ }H_{0}^{1}\cap W^{1,p}\left(
\Omega \right) ;\quad u_{1m}\longrightarrow u_{1}\quad \text{in \ \ }H\cap
L^{p}\left( \Omega \right) \text{ at }\ m\nearrow \infty .
\end{equation*}

Thus we obtain the following problem 
\begin{equation}
\frac{d^{2}}{dt^{2}}\left\langle u_{m},w_{j}\right\rangle =\left\langle
F\left( u_{m}\right) ,\Delta w_{j}\right\rangle +\left\langle g\left(
u_{m}\right) ,w_{j}\right\rangle ,\quad j=1,2,...,m  \label{3.3}
\end{equation}%
\begin{equation*}
\left\langle u_{m}\left( t,x\right) ,w_{j}\right\rangle \left\vert \
_{t=0}\right. =\left\langle u_{0m}\left( x\right) ,w_{j}\right\rangle ,\ 
\frac{d}{dt}\left\langle u_{m}\left( t,x\right) ,w_{j}\right\rangle
\left\vert \ _{t=0}\right. =\left\langle u_{1m}\left( x\right)
,w_{j}\right\rangle
\end{equation*}%
that solvable on $\left( 0,T\right] $ for any $m=1,2,...$ and $T>0$ by
virtue of estimates (\ref{2.5})$.$

Consequently with use of the known procedure (\cite{li}, \cite{z}, \cite{s}%
]) we obtain, $\nabla v_{mt}\in C^{0}\left( 0,T;H\left( \Omega \right)
\right) ,$ $\Delta v_{m}\in C^{0}\left( 0,T;L^{p}\left( \Omega \right)
\right) $ and $u_{m}\in C^{0}\left( 0,T;L^{p}\left( \Omega \right) \right) $%
, $\ u_{mt}\in C^{0}\left( 0,T;H^{-1}\left( \Omega \right) \right) $,
moreover they are contained in the bounded subset of these spaces. Thus from
(\ref{3.3}) we get 
\begin{equation*}
u_{mtt}\in C^{0}\left( 0,T;\left( W^{2,p}\left( \Omega \right) \cap
W_{0}^{1,p}\left( \Omega \right) \right) ^{\ast }+H^{-1}\left( \Omega
\right) \right) .
\end{equation*}%
So for the sequence of the approximate solutons we have: $\left\{
u_{m}\right\} _{m=1}^{\infty }$ is contained in the bounded subset of the
space 
\begin{equation*}
C^{0}\left( 0,T;L^{p}\left( \Omega \right) \right) \cap C^{2}\left(
0,T;W^{-2,q}\left( \Omega \right) +H^{-1}\left( \Omega \right) \right)
\end{equation*}%
and the sequence $\left\{ v_{m}\right\} _{m=1}^{\infty }$ is contained in
the bounded subset of the space 
\begin{equation*}
C^{0}\left( 0,T;W^{2,p}\left( \Omega \right) \cap H_{0}^{1}\left( \Omega
\right) \right) \cap C^{1}\left( 0,T;H_{0}^{1}\left( \Omega \right) \right)
\cap C^{2}\left( 0,T;L^{q}\left( \Omega \right) \right) .
\end{equation*}%
Then the sequence $\left\{ v_{m}\right\} _{m=1}^{\infty }$ has a precompact
subset in the space 
\begin{equation*}
C^{1}\left( 0,T;\left[ W^{2,p}\left( \Omega \right) ,L^{q}\left( \Omega
\right) \right] _{\frac{1}{2}}\right)
\end{equation*}
by virtue of the known interpolation theory (see, \cite{t}), and
consequently in the space $C^{1}\left( 0,T;H_{0}^{1}\left( \Omega \right)
\right) $ as the imbedding $\left[ W^{2,p}\left( \Omega \right) ,L^{q}\left(
\Omega \right) \right] _{\frac{1}{2}}\subseteq H^{1}\left( \Omega \right) $
holds.

Thus for us is remained to show the following: if the sequence 
\begin{equation*}
\left\{ u_{m}\right\} _{m=1}^{\infty }\subset C^{0}\left( 0,T;L^{p}\left(
\Omega \right) \right) \cap C^{2}\left( 0,T;W^{-2,q}\left( \Omega \right)
+H^{-1}\left( \Omega \right) \right)
\end{equation*}%
is weakly converging to $u$ in this space and $\left\{ F\left( u_{m}\right)
\right\} _{m=1}^{\infty }$ and $\left\{ g\left( u_{m}\right) \right\}
_{m=1}^{\infty }$ have an weakly converging subsequence to $\eta $ in $H$
and to $\theta $ in $L^{q}\left( \Omega \right) $ ($q=\frac{p}{p-1}$)
respectively, for a. e. $t\in \left( 0,T\right) $ then $\eta =F\left(
u\right) $ and $\theta =g\left( u\right) $. (Here and in what follows for
brevity we don't changing of indexes of subsequences.)

In the beginning we will show the equation $\theta =g\left( u\right) $. Let
the sequence $\left\{ u_{m}\right\} _{m=1}^{\infty }$ is such as above
mentioned and $-\Delta v_{m}=u_{m}$. Then in condition (\ref{3.1}) for the
operator 
\begin{equation*}
g:C^{0}\left( 0,T;L^{p}\left( \Omega \right) \right) \subset C^{0}\left(
0,T;H\right) \longrightarrow C^{0}\left( 0,T;H\right)
\end{equation*}%
we have 
\begin{equation*}
\left\langle g\left( u_{m}\right) ,w_{j}\right\rangle \longrightarrow
\left\langle \theta ,w_{j}\right\rangle \text{ for }\forall j:\text{ }%
j=1,2,...\text{\ }
\end{equation*}%
and also 
\begin{equation*}
\left\langle g\left( u_{m}\right) ,z\right\rangle \longrightarrow
\left\langle \theta ,z\right\rangle \text{ \ for }\forall z\in W^{2,p}\left(
\Omega \right) \subset H\left( \Omega \right) .
\end{equation*}%
Therefore we consider the expression $\left\langle g\left( u_{m}\right)
,v_{m}\right\rangle $ under the assumption that $u_{m}\rightharpoonup u$ in $%
L^{p}\left( \Omega \right) \subset H$ and $v_{m}\longrightarrow v$ in $%
H^{1}\left( \Omega \right) $ and $g\left( u_{m}\right) \rightharpoonup
\theta $ in the corresponding spaces. In order to prove that $\left\langle
g\left( u_{m}\right) ,v_{m}\right\rangle $ is the Cauchy sequence we carry
out the following estimations 
\begin{equation*}
\left\vert \left\langle g\left( u_{m}\right) ,v_{m}\right\rangle
-\left\langle g\left( u_{m+k}\right) ,v_{m+k}\right\rangle \right\vert \leq
\left\vert \left\langle g\left( u_{m}\right) -g\left( u_{m+k}\right)
,v_{m}\right\rangle \right\vert +
\end{equation*}%
\begin{equation*}
\left\vert \left\langle g\left( u_{m+k}\right) ,v_{m}-v_{m+k}\right\rangle
\right\vert \leq \left\langle \left\vert g\left( u_{m}\right) -g\left(
u_{m+k}\right) \right\vert ,\left\vert v_{m}\right\vert \right\rangle +
\end{equation*}%
\begin{equation}
\left\vert \left\langle g\left( u_{m+k}\right) ,v_{m}-v_{m+k}\right\rangle
\right\vert \leq d_{0}\left\langle \left\vert u_{m}-u_{m+k}\right\vert
,\left\vert v_{m}\right\vert \right\rangle +\left\vert \left\langle g\left(
u_{m+k}\right) ,v_{m}-v_{m+k}\right\rangle \right\vert  \label{3.4}
\end{equation}%
that shows the correctnes of this statement as the right side converge to
zero with respect to $m\nearrow \infty $. If to take account of the above
assumpsion we can conduct the estimation of such type (\ref{3.4}) for the
expression $\left\vert \left\langle g\left( u_{m}\right) ,v_{m}\right\rangle
-\left\langle g\left( u\right) ,v\right\rangle \right\vert $, as $g\left(
u\right) $ is defined,\ then we obtain that equation $\theta =g\left(
u\right) $ holds, i.e. $g\left( u_{m}\right) \rightharpoonup g\left(
u\right) $ in $H$.

In order to show the equation $\eta =F\left( u\right) $ we will use the
monotonicity condition of F, i. e. for any $v,w\in C^{0}\left(
0,T;W^{2,p}\left( \Omega \right) \right) \cap C^{2}\left( 0,T;L^{p}\left(
\Omega \right) \right) $ occurs the following inequation 
\begin{equation*}
\left\langle -\Delta F\left( -\Delta v\right) +\Delta F\left( -\Delta 
\widetilde{v}\right) ,v-\widetilde{v}\right\rangle \geq 0
\end{equation*}%
and if rewrite it for $u_{m}=-\Delta v_{m}$ and $\widetilde{u}=-\Delta 
\widetilde{v}$ then we have 
\begin{equation*}
\left\langle \left( F\left( u_{m}\right) -F\left( \widetilde{u}\right)
\right) ,u_{m}-\widetilde{u}\right\rangle \geq 0.
\end{equation*}

It is not difficult to see that the following convergence takes place 
\begin{equation*}
\frac{d}{dt}\left\langle u_{mt},w_{j}\right\rangle -\left\langle \Delta
F\left( u_{m}\right) ,w_{j}\right\rangle -\left\langle g\left( u_{m}\right)
,w_{j}\right\rangle \longrightarrow \frac{d}{dt}\left\langle
u_{t},w_{j}\right\rangle -\left\langle \Delta \eta ,w_{j}\right\rangle
-\left\langle \theta ,w_{j}\right\rangle ,\quad \forall w_{j}
\end{equation*}%
then 
\begin{equation}
\frac{d^{2}}{dt^{2}}\left\langle u,w\right\rangle -\left\langle \Delta \eta
,w\right\rangle =\left\langle g\left( u\right) ,w\right\rangle ,\quad
\forall w\in H_{0}^{1}\cap W^{2,p}\left( \Omega \right)  \label{3.5}
\end{equation}%
for a. e. $t\in \left( 0,T\right) $ by virtue of the obtained above equation 
$\theta =g\left( u\right) $, consequently 
\begin{equation*}
u_{tt}-\Delta \eta =g\left( u\right) ,\quad \text{in the sense of }%
H^{-1}+W^{-2,q}\left( \Omega \right)
\end{equation*}%
for a. e. $t\in \left( 0,T\right) $.

Let us apply monotonicity of $F$ 
\begin{equation*}
0\leq \left\langle F\left( u_{m}\right) -F\left( \widetilde{u}\right) ,u_{m}-%
\widetilde{u}\right\rangle =-\left\langle \Delta F\left( u_{m}\right)
+\Delta F\left( \widetilde{u}\right) ,v_{m}-\widetilde{v}\right\rangle =
\end{equation*}%
\begin{equation*}
-\left\langle \Delta F\left( u_{m}\right) ,v_{m}\right\rangle +\left\langle
\Delta F\left( u_{m}\right) ,\widetilde{v}\right\rangle +\left\langle \Delta
F\left( \widetilde{u}\right) ,v_{m}-\widetilde{v}\right\rangle =
\end{equation*}%
(here $\widetilde{u}=-\Delta \widetilde{v}$, $\widetilde{v}\in H_{0}^{1}\cap
W^{2,p}\left( \Omega \right) $) for that use equation (\ref{3.3}) we have 
\begin{equation*}
\left\langle -g\left( u_{m}\right) +\frac{\partial ^{2}}{\partial t^{2}}%
u_{m},\widetilde{v}\right\rangle -\left\langle \Delta F\left( u_{m}\right)
,v_{m}\right\rangle +\left\langle \Delta F\left( \widetilde{u}\right)
,v_{m}-v\right\rangle =
\end{equation*}%
\begin{equation*}
-\left\langle g\left( u_{m}\right) ,\widetilde{v}\right\rangle +\frac{d^{2}}{%
dt^{2}}\left\langle u_{m},\widetilde{v}\right\rangle +\left\langle F\left(
u_{m}\right) ,u_{m}\right\rangle +\left\langle \Delta F\left( \widetilde{u}%
\right) ,v_{m}-\widetilde{v}\right\rangle \Longrightarrow
\end{equation*}%
from here\ we get 
\begin{equation*}
0\leq -\left\langle g\left( u\right) ,\widetilde{v}\right\rangle +\frac{d^{2}%
}{dt^{2}}\left\langle u,\widetilde{v}\right\rangle +\left\langle F\left(
u_{m}\right) ,u_{m}\right\rangle +\left\langle \Delta F\left( \widetilde{u}%
\right) ,v-\widetilde{v}\right\rangle
\end{equation*}%
by pass to the limit with respect to $m:m\nearrow \infty $ and if to take
account the following known inequation 
\begin{equation*}
\underset{\Omega }{\int }\lim \inf \left( F\left( u_{m}\right) u_{m}\right)
dx\leq \left\langle \eta ,u\right\rangle ,
\end{equation*}%
by the Fatou's lemma, more exactly 
\begin{equation*}
\underset{\Omega }{\int }\lim \inf \left( F\left( -\Delta v_{m}\right)
\left( -\Delta v_{m}\right) \right) dx\leq \left\langle \eta ,u\right\rangle
\end{equation*}%
as $\left\langle -\Delta F\left( u_{m}\right) ,v_{m}\right\rangle
=\left\langle F\left( -\Delta v_{m}\right) ,-\Delta v_{m}\right\rangle $.
Then with use of this inequation and (\ref{3.5}) we get 
\begin{equation*}
0\leq -\left\langle g\left( u\right) ,\widetilde{v}\right\rangle +\frac{d^{2}%
}{ds^{2}}\left\langle \Delta v,\widetilde{v}\right\rangle -\left\langle
\Delta \eta ,v\right\rangle +\left\langle \Delta F\left( \widetilde{u}%
\right) ,v-\widetilde{v}\right\rangle =
\end{equation*}%
\begin{equation*}
\left\langle -\Delta \eta ,v-\widetilde{v}\right\rangle -\left\langle
-\Delta F\left( \widetilde{u}\right) ,v-\widetilde{v}\right\rangle
=\left\langle \eta -F\left( \widetilde{u}\right) ,u-\widetilde{u}%
\right\rangle .
\end{equation*}%
Hence we obtain the equation $\eta =F\left( u\right) $ by virtue of
arbitrariness of $\widetilde{u}=-\Delta \widetilde{v}$.

Now we will show that the function $u\left( t,x\right) $ satisfies of the
initial conditions and for this we will consider the following equation 
\begin{equation*}
\left\langle u_{mt},v_{m}\right\rangle \left( t\right) =\underset{0}{\overset%
{t}{\int }}\left\langle u_{mss},v_{m}\right\rangle ds+\underset{0}{\overset{t%
}{\int }}\left\langle u_{ms},v_{ms}\right\rangle ds+\left\langle
u_{1m},v_{0m}\right\rangle
\end{equation*}%
for $t\in \left( 0,T\right) $ and $u_{m}=-\Delta v_{m}$, that is equivalent
to the equation 
\begin{equation}
\left\langle \nabla v_{mt},\nabla v_{m}\right\rangle \left( t\right) =%
\underset{0}{\overset{t}{\int }}\left\langle v_{mss},u_{m}\right\rangle ds+%
\underset{0}{\overset{t}{\int }}\left\langle \nabla v_{ms},\nabla
v_{ms}\right\rangle ds+\left\langle \nabla v_{1m},\nabla v_{0m}\right\rangle
.  \label{3.6}
\end{equation}%
From obtained a priory estimations follow the boundedness of the right side
of (\ref{3.6}), consequently we get the boundedness of the left side of (\ref%
{3.6}) any $t\in \left( 0,T\right) $. Thus one can pass to the limit by $%
t\longrightarrow 0$ by virtue of the a priory estimations. Really as $%
\left\{ v_{m}\right\} _{m=1}^{\infty }\in $ $C^{0}\left( 0,T;W^{2,p}\left(
\Omega \right) \right) \cap C^{2}\left( 0,T;L^{q}\left( \Omega \right)
\right) $ and bounded in this space we get: the right side is bounded as all
terms in the left side are bounded in respective spaces, therefore one can
pass to limit with respect to $m$ as here $v_{mt}$ are continous with
respect to $t$ for any $m$ then $v_{mt}$ strongly converges to $v_{t}$ in $H$
and $\Delta v_{m}$ weakly converges to $\Delta v$ in $L^{p}\left( \Omega
\right) $.

Consequently is proved the following result.

\begin{theorem}
Let $u_{0}\in $\ $H_{0}^{1}\cap W^{1,p}\left( \Omega \right) $ , $u_{1}\in $%
\ $L^{p}\left( \Omega \right) $ and that there are functions $v_{0}\in $\ $%
H_{0}^{1}\cap W^{2,p}\left( \Omega \right) $ , $v_{1}\in $\ $H_{0}^{1}\cap
W^{1,p}\left( \Omega \right) $ such that $-\Delta v_{k}=u_{k}$, $k=0,1$. Let 
$f:R\longrightarrow R$ \ and $g:R\longrightarrow R$ are a continuous
functions such that $F\left( u\right) =\underset{0}{\overset{u}{\int }}%
f\left( r\right) dr$ is a\ monotone operator and satisfies the following
inequalities 
\begin{equation*}
\left\vert F\left( r\right) \right\vert \leq a_{0}\left\vert r\right\vert
^{p-1}+a_{1}\left\vert r\right\vert ;\quad F\left( r\right) \cdot r\geq \
b_{0}\left\vert r\right\vert ^{p}+b_{1}r^{2}
\end{equation*}%
for $r\in R$, and $g$ satisfies inequation (\ref{3.1}) for $r,s\in R$, where 
$a_{0},b_{0},d_{0}>0$, $a_{1},b_{1}\geq 0$ and $p>1$ are a numbers.

Then problem (\ref{2.1})-(\ref{2.2}) possess, at least, one weak solution $%
u\left( t,x\right) $ in the sense of Definition\ref{D-1} that belongs to the
space $C^{0}\left( 0,T;L^{p}\left( \Omega \right) \right) \cap C^{2}\left(
0,T;W^{-2,q}\left( \Omega \right) +H^{-1}\left( \Omega \right) \right) $ for
every fixed number $T\in \left( 0,\infty \right) $.
\end{theorem}

\begin{remark}
It should be noted that by using (\ref{2.4}) one can reformulate of the
considered problem in the following form: let $g\left( u\right) \equiv
g\left( t,x\right) $ is given function 
\begin{equation*}
-\Delta \left( v_{tt}+F\left( -\Delta v\right) \right) =g\left( t,x\right)
\equiv -\Delta \widetilde{g},\quad \left( t,x\right) \in \left( 0,T\right)
\times \Omega ,
\end{equation*}%
\begin{equation*}
-\Delta v\left( 0,x\right) =u_{0}\left( x\right) =-\Delta v_{0}\left(
x\right) ,\ 
\end{equation*}%
\begin{equation*}
-\Delta v_{t}\left( 0,x\right) =u_{1}\left( x\right) =-\Delta v_{1}\left(
x\right) ,\ \Delta v\left\vert \ _{\left( 0,T\right) \times \partial \Omega
}\right. =0.
\end{equation*}%
In the other words we get 
\begin{equation}
-\Delta \left( v_{tt}+F\left( -\Delta v\right) -\widetilde{g}\right)
=0,\quad \left( t,x\right) \in \left( 0,T\right) \times \Omega ,  \label{3.7}
\end{equation}%
hence one can obtain the following equivalent problem if $F$ is the
homogeneous operator 
\begin{equation*}
v_{tt}+F\left( -\Delta v\right) =\widetilde{g},\quad \left( t,x\right) \in
\left( 0,T\right) \times \Omega ,
\end{equation*}%
\begin{equation*}
v\left( 0,x\right) =v_{0}\left( x\right) ,\ v_{t}\left( 0,x\right)
=v_{1}\left( x\right) ,\ v\left\vert \ _{\left( 0,T\right) \times \partial
\Omega }\right. =-\Delta v\left\vert \ _{\left( 0,T\right) \times \partial
\Omega }\right. =0
\end{equation*}%
since if equation (\ref{3.7}) possess a solution then the expression $%
v_{tt}+F\left( -\Delta v\right) -\widetilde{g}$ is a harmonic function for
any $t$ and also satisfies the homogeneous boundary condition. In this case
we get, that the considered problem is equivalent to the problem 
\begin{equation*}
v_{tt}+F\left( -\Delta v\right) =\widetilde{g}\left( t,x\right) \text{ with
the mixed conditions as above.}
\end{equation*}
\end{remark}

\section{Behaviour of the solution of the problem (\protect\ref{2.1})-(%
\protect\ref{2.2})}

Now we introduce the function $E(t)=\Vert \nabla w\Vert _{H}^{2}(t)$ and
consider this function on the solution of problem (\ref{2.1})-(\ref{2.2})
and assume $g$ satisfies inequation $\left\vert g\left( r\right) \right\vert
^{2}\leq d_{1}\Phi \left( r\right) $ for any $r\in R$.

So we will study\ the problem 
\begin{equation*}
u_{tt}-\Delta F\left( u\right) =g\left( u\right) ,\quad \left( t,x\right)
\in R_{+}\times \Omega ,
\end{equation*}%
\begin{equation*}
u\left( 0,x\right) =u_{0}\left( x\right) ,\ u_{t}\left( 0,x\right)
=u_{1}\left( x\right) ,\ u\left\vert \ _{R_{+}\times \partial \Omega
}\right. =0
\end{equation*}%
for which behaving as above we get the equation 
\begin{equation*}
\left\Vert \nabla v_{t}\right\Vert _{2}^{2}\left( t\right) +2\Phi \left(
-\Delta v\right) \left( t\right) =\left\Vert \nabla v_{1}\right\Vert
_{2}^{2}+2\Phi \left( -\Delta v_{0}\right) +2\left\langle g\left( u\right)
,v_{t}\right\rangle
\end{equation*}%
(it not is difficult to see that if $g\left( u\right) \equiv 0$ then the
energy functional remain constant for $\forall t>0$, i. e. the energy
functional is independent of $t>0$) 
\begin{equation*}
\left\Vert \nabla v_{t}\right\Vert _{2}^{2}\left( t\right) +2\Phi \left(
u\right) \left( t\right) \leq \left\Vert \nabla v_{1}\right\Vert
_{2}^{2}+2\Phi \left( u_{0}\right) +\underset{0}{\overset{t}{\int }}\left[
\left\Vert v_{s}\right\Vert _{2}^{2}+\left\Vert g\left( u\right) \right\Vert
_{2}^{2}\right] \left( s\right) ds
\end{equation*}%
using the condition on $g\left( u\right) $ we have 
\begin{equation*}
\left\Vert \nabla v_{t}\right\Vert _{2}^{2}\left( t\right) +2\Phi \left(
u\right) \left( t\right) \leq \left\Vert \nabla v_{1}\right\Vert
_{2}^{2}+2\Phi \left( u_{0}\right) +\widetilde{d}\underset{0}{\overset{t}{%
\int }}\left[ \left\Vert \nabla v_{s}\right\Vert _{2}^{2}+2\Phi \left(
u\right) \right] \left( s\right) ds.
\end{equation*}%
Hence follows 
\begin{equation}
\left\Vert \nabla v_{t}\right\Vert _{2}^{2}\left( t\right) \leq \frac{1}{%
\widetilde{d}}\left[ e^{\widetilde{d}t}\left( 1+\widetilde{d}\right) -1%
\right] \left( \left\Vert \nabla v_{1}\right\Vert _{2}^{2}+2\Phi \left(
u_{0}\right) \right) -2\Phi \left( u\right) \left( t\right) .  \label{4.1}
\end{equation}

For the functional $E\left( t\right) =\left\Vert \nabla v\right\Vert
_{2}^{2}\left( t\right) $ we have 
\begin{equation*}
E`\left( t\right) =2\left\langle \nabla v_{t},\nabla v\right\rangle \leq
\left\Vert \nabla v_{t}\right\Vert _{2}^{2}\left( t\right) +\left\Vert
\nabla v\right\Vert _{2}^{2}\left( t\right)
\end{equation*}%
using here inequation (\ref{4.1}) 
\begin{equation*}
E`\left( t\right) \leq E\left( t\right) -2\Phi \left( u\right) \left(
t\right) +\frac{1}{\widetilde{d}}\left[ e^{\widetilde{d}t}\left( 1+%
\widetilde{d}\right) -1\right] \left( \left\Vert \nabla v_{1}\right\Vert
_{2}^{2}+2\Phi \left( u_{0}\right) \right)
\end{equation*}%
and the condition on $F$ (consequently, on $\Phi $) 
\begin{equation*}
E`\left( t\right) \leq E\left( t\right) -c\left\Vert -\Delta v\right\Vert
^{p}\left( t\right) +\frac{1}{\widetilde{d}}\left[ e^{\widetilde{d}t}\left(
1+\widetilde{d}\right) -1\right] \left( \left\Vert \nabla v_{1}\right\Vert
_{2}^{2}+2\Phi \left( u_{0}\right) \right) \leq
\end{equation*}%
\begin{equation*}
E\left( t\right) -cE^{\frac{p}{2}}\left( t\right) +\frac{1}{\widetilde{d}}%
\left[ e^{\widetilde{d}t}\left( 1+\widetilde{d}\right) -1\right] \left(
\left\Vert \nabla v_{1}\right\Vert _{2}^{2}\left( 0\right) +2\Phi \left(
-\Delta v_{0}\right) \right) \Longrightarrow
\end{equation*}%
and at last we get 
\begin{equation*}
E`\left( t\right) \leq E\left( t\right) -cE^{\frac{p}{2}}\left( t\right)
+C_{1}\left( v_{0},v_{1}\right) e^{\widetilde{d}t}-C_{2}\left(
v_{0},v_{1}\right)
\end{equation*}%
by virtue of the condition $\Phi \left( r\right) \geq c_{0}\left\vert
r\right\vert ^{p}$ and of the continuity of embeddings $L^{p}\left( \Omega
\right) \subset L^{2}\left( \Omega \right) ,$ \ $W^{2,p}\left( \Omega
\right) \subset \subset W^{1,p}\left( \Omega \right) $, where $C_{j}\left(
v_{0},v_{1}\right) >0$ ( $j=1,2$) are constants.

So we have the Cauchy problem for differential inequality 
\begin{equation}
y`\left( t\right) \leq y\left( t\right) -cy^{r}\left( t\right) +C_{1}e^{%
\widetilde{d}t}-C_{2},\quad y\left( 0\right) =\left\Vert \nabla
v_{0}\right\Vert _{2}^{2}.  \label{4.2}
\end{equation}%
One can replace problem (\ref{4.2}) with the following problem in order to
investigate of the behaviour of the solution of considered problem 
\begin{equation*}
y`\left( t\right) \leq y\left( t\right) -cy^{r}\left( t\right) +C_{1}e^{%
\widetilde{d}T}-C_{2},\quad y\left( 0\right) =\left\Vert \nabla
v_{0}\right\Vert _{2}^{2}
\end{equation*}%
as $\widetilde{d}>0$. Inequation (\ref{4.2}) one can rewrite in the form 
\begin{equation*}
\left( y\left( t\right) +lC\left( v_{0},v_{1}\right) \right) 
{\acute{}}%
\leq y\left( t\right) +lC\left( v_{0},v_{1}\right) -\varepsilon \left[
y\left( t\right) +lC\left( v_{0},v_{1}\right) \right] ^{r},
\end{equation*}%
where $l>1$ is a number and $\varepsilon =\varepsilon \left( c,C,l,r\right)
>0$ is sufficiently small number and $C=C\left( \widetilde{d}%
,T,C_{1},C_{2}\right) $. Then solving this problem we get 
\begin{equation*}
y\left( t\right) +lC\left( v_{0},v_{1}\right) \leq \left[ e^{\left(
1-r\right) t}\left( \nabla y_{0}+lC\left( v_{0},v_{1}\right) \right)
^{1-r}+\varepsilon \left( 1-e^{\left( 1-r\right) t}\right) \right] ^{\frac{1%
}{1-r}}
\end{equation*}%
or 
\begin{equation*}
E\left( t\right) \leq \left[ e^{\left( 1-r\right) t}\left( \left\Vert \nabla
v_{0}\right\Vert _{H}^{2}+lC\left( v_{0},v_{1}\right) \right)
^{1-r}+\varepsilon \left( 1-e^{\left( 1-r\right) t}\right) \right] ^{\frac{1%
}{1-r}}-lC\left( v_{0},v_{1}\right)
\end{equation*}%
\begin{equation}
\Vert \nabla v\Vert _{H}^{2}(t)\leq \frac{e^{t}\left( \left\Vert \nabla
v_{0}\right\Vert _{H}^{2}+lC\left( v_{0},v_{1}\right) \right) }{\left[
1+\varepsilon \left( \left\Vert \nabla v_{0}\right\Vert _{H}^{2}+lC\left(
v_{0},v_{1}\right) \right) ^{r-1}\left( e^{\left( r-1\right) t}-1\right) %
\right] ^{\frac{1}{r-1}}}-lC\left( v_{0},v_{1}\right) .  \label{4.3}
\end{equation}%
here the right side is greater than zero, because $\varepsilon \leq \frac{l-1%
}{l^{r}C^{r}}$ and $2r=p>2$. It is necessary to note here the dependence on $%
T$ of the behaviour of the solution is essentially that follows from the
received last problem.

Thus is proved the result

\begin{lemma}
Let $u_{0}\in $\ $H_{0}^{1}\cap W^{1,p}\left( \Omega \right) $ , $u_{1}\in $%
\ $L^{p}\left( \Omega \right) $ and that there are functions $v_{0}\in $\ $%
H_{0}^{1}\cap W^{2,p}\left( \Omega \right) $ , $v_{1}\in $\ $H_{0}^{1}\cap
W^{1,p}\left( \Omega \right) $ such that $-\Delta v_{k}=u_{k}$, $k=0,1$.
Then the function $v(t,x)$, defined by the solution $u\left( t,x\right) $ of
problem (\ref{2.1})-(\ref{2.2}), for any $t\in \left( ),T\right) $ belong in
ball $B_{R_{T}}^{H_{0}^{1}\cap W^{1,p}\left( \Omega \right) }\left( 0\right)
\subset H_{0}^{1}\cap W^{1,p}\left( \Omega \right) $ depending from the
initial values $\left( u_{0},u_{1}\right) \in H_{0}^{1}\cap W^{1,p}\left(
\Omega \right) \times L^{p}\left( \Omega \right) $, here $R_{T}=R_{T}\left(
u_{0},u_{1},p,T\right) >0$. \ 
\end{lemma}

\end{document}